\documentstyle[twoside,psfig,ijmpd]{article}

\def\b{\begin{equation}}
\def\e{\end{equation}}
\def\pa{\partial}
\def\l{\left}
\def\r{\right}

\begin{document}
\setlength{\textheight}{7.7truein}  

\runninghead{S. Shankaranarayanan and T. Padmanabhan}
{Hypothesis of path integral duality: Applications to QED}

\normalsize
\textlineskip
\thispagestyle{empty}
\setcounter{page}{1}
\copyrightheading{}                     
\vspace*{0.88truein}
\fpage{1}


\centerline{Hypothesis of path integral duality: Applications to QED}
\vspace*{0.37truein}
\centerline{\footnotesize S.~Shankaranarayanan\footnote{Electronic 
address:~shanki@iucaa.ernet.in}} 
\vspace*{6pt}
\centerline{\footnotesize T.~Padmanabhan\footnote{Electronic 
address:~paddy@iucaa.ernet.in}}
\baselineskip=12pt
\centerline{\footnotesize\it IUCAA, Post Bag 4, Ganeshkhind}
\baselineskip=10pt
\centerline{\footnotesize\it Pune 411 007, INDIA.}
\vspace*{0.225truein}

\publisher{(received date)}{(revised date)}

\vspace*{0.21truein}

\abstracts{ We use the modified propagator for quantum field  based on a 
``principle of 
path integral duality" proposed earlier in a paper by Padmanabhan to 
investigate several results in QED. This procedure modifies the Feynman 
propagator by the introduction of a fundamental length scale. We use this 
modified propagator for the Dirac particles to evaluate the first order 
radiative corrections in QED. We find that the extra factor of the modified 
propagator acts like a regulator at the Planck scales thereby removing the 
divergences that otherwise appear in the conventional radiative correction
 calculations of QED. We find that:(i) all the three renormalization factors 
$Z_1$, $Z_2$, and $Z_3$ pick up finite corrections and (ii) the modified
 propagator breaks the gauge invariance at a very small level of 
${\mathcal{O}}(10^{-45})$. The implications of this result to generation of 
the primordial seed magnetic fields are discussed.}{}{} 

\vspace*{1pt}\textlineskip      
\section{Introduction}    
\vspace*{-0.5pt}
\noindent
	 The space-time structure at Planck scales $L_P \equiv 
(G\hbar/c^3)^{1/2}$ will be drastically affected by the quantum gravitational 
effects and it is generally believed that $L_P$ acts as a physical cutoff for 
space-time intervals. The two main approaches to quantum gravity, Superstring 
theory and Loop quantum gravity, incorporates the fundamental length scale by 
considering extended structures, rather than point particles, as fundamental 
blocks.  

	The existence of a fundamental length implies that processes 
involving energies higher than Planck energies will be suppressed, and the 
ultraviolet behavior of the theory will be improved. This could arise naturally in String theories; several other models also incorporate a Planck length 
cut-off in a suitable manner to improve the ultra violet behaviour of the 
theory \cite{ahluwalia}. One direct consequence of such improved behavior 
will be that the Feynman propagator(in momentum space) will acquire damping 
factor for energies larger than Planck energy. However, this propagator -- 
which arises in the standard formulation of quantum field theory -- does not 
take into account of the existence of any 
fundamental length in the space-time. On the other hand, such a 
fundamental length scale was introduced into the Feynman propagator in a 
Lorentz invariant manner by invoking the ``principle of path-integral 
duality" \cite{paddy1}. According to this postulate, the weightage given for 
a path in the path integral should be invariant under the transformation 
${\mathcal{R}} \rightarrow L_P^2/\mathcal{R}$, where $\mathcal{R}$ is the 
length of the path and the fundamental length scale $L_P$ is assumed to be of 
the order of the Planck length $(G \hbar/c^3)^{1/2}$.[In this paper, when we 
say that the fundamental length is $L_P$, we actually mean that it is of 
${\mathcal{O}} (L_P)$.] Padmanabhan \cite{paddy1} has shown by rigorous 
evaluation of the path integral by lattice techniques that the effect 
of the duality principle is to modify the weightage given to a path of proper 
time $s$ from $\exp(-i m^2s)$ to $\exp[-i (m^2 s - L_P^2/s)]$, where $m$ is 
associated with the mass of the particle. For example, the Feynman propagator 
for a free scalar field of mass $m$, propagating in the flat $(3 + 1)$ 
space-time, in Schwinger's proper time formalism, is described by the 
integral
\b 
G_F(x,x') = \frac{1}{(4 \pi i)^2}\int_{0}^{\infty} \frac{ds}{s^2} 
\exp(-i m^2 s) \exp[i(x - x')^2/4s]. 
\e
\noindent The duality principle modifies the Feynman propagator to the form 
\begin{eqnarray} {\label{modfey}}
G_F^P(x,x')&=&\frac{1}{(4 \pi i)^2}\int_{0}^{\infty}  \frac{ds}{s^2} 
\exp(-i m^2 s) \exp[i((x-x')^2 + L_P^2)/4s] \nonumber \\
&=&\frac{m}{4 \pi^2} \frac{K_1(i m \sqrt{x^2 - L_P^2 -i\epsilon})}
{\sqrt{x^2 - L_P^2 -i\epsilon}}, 
\end{eqnarray}
\noindent where $K_1(z)$ is the modified Bessel function of order 1.
[The metric signature we follow is $(+,-,-,-)$.] In momentum space, the 
modified propagator is  
\b {\label {maineq}}
G_F^P(p) =  -i \int_0^{\infty} dz \exp(i L_P^2/4z + i(p^2 - m^2 +i\epsilon)z).
\e

	The presence of a fundamental length scale is a feature that is 
expected to arise in a quantum theory of gravity. Hence, the modification 
of the weightage factor as mentioned above can be interpreted as being 
equivalent to introducing quantum gravitational corrections into standard 
field theory. The ultraviolet divergences in quantum field theory arise from 
the singularities of the propagator functions on the light cone, and a 
smearing out of the light cone due to the quantum gravitational corrections, 
using the principle of ``path integral duality'',  will lead to the 
suppression of these divergences.
  
       In an earlier work \cite{paddy2}, the implications of the modified 
propagator to certain conventional non-perturbative quantum field theoretic 
results were discussed in detail. It was found that the essential feature of 
this prescription of path integral duality is to provide an ultra-violet 
cutoff at the Planck energy scales, thereby obtaining a Lorentz invariant 
finite results. Encouraged by this fact, we would like to estimate the 
renormalization factors in QED, and other  radiative correction terms which 
cannot be obtained in the conventional QED calculations.  
         
	The standard definition of Feynman propagator for Dirac particles is 
\b
S(x) = -(-i \gamma^{\mu} \partial_{\mu} + m) G_F(x),
\e 
\noindent where, $G_F(x)$ is the usual Feynman propagator for the scalar 
particles. In our analysis of evaluating quantum gravitational corrections to 
the standard QED calculations, we assume that the effect of ``principle of 
path integral duality" on the Dirac propagator is defined as 
\b
S^P(x) = -(-i \gamma^{\mu} \partial_{\mu} + m) G_F^P(x).
\e
\noindent Hence, the effect of summing over the quantum fluctuations of the 
space-time structure, in the low energy scales, can be realized in any field 
theoretic calculations where the propagator of the scalar particles appear 
explicitly. 

	 There is a similar approach in the literature by Ohanian 
\cite{ohanian1}, who had used a smeared propagator which is Poincare invariant,
 to calculate the radiative corrections in QED. The modified Feynman 
propagator in our case is obtained by rigorous evaluation and the duality 
principle introduces the fundamental length scale in a Lorentz invariant 
manner. We also comment on the differences between our approaches and results 
at the appropriate sections below.
	
	Before proceeding to the technical aspects, it is necessary to
outline certain conceptual issues related to this approach. In the conventional 
approaches to quantizing a field based on some classical Lagrangian, one will
invariably obtain quantum corrections to the classical Lagrangian. When these 
corrections are local in the configuration space and contain terms which are
 of the same form as those in the classical Lagrangian, it is necessary to 
absorb them into the parameters of the original Lagrangian. This process of 
renormalization, has --- {\it a priori} --- nothing to do with 
divergences. Usually, however, the quantum corrections to the theory --- 
calculated by perturbative methods --- lead to divergent expressions. 
If the new (divergent) terms are not of the form of the original terms, 
then the theory cannot be interpreted perturbatively. However, if the 
divergent terms have the same structure as the terms in the original 
Lagrangian, it is possible to use the procedure of renormalization 
(which, {\it a priori}, has nothing to do with divergences) to give 
meaning to the theory. To do this properly, it is necessary to first evolve 
a procedure (called regularization) which allows the divergent expressions to 
be recast as the limit of some finite quantities. After performing the 
renormalization subtractions, one is free to take the required limit, leading 
to finite corrections. 

	The above approach gets modified in two essential aspects when the 
modified propagator is used. 
Firstly, when quantum gravitational corrections lead to a cutoff, the quantum 
corrections will not have any divergent terms; that is, the regularization is 
now built into the theory with Planck length acting as regulator. But 
renormalization of the theory is still needed and the physical and bare 
coupling constants will differ by a finite amount. Secondly, {\it the
 regularization procedure is now fixed by our ansatz}. This is important 
because, it is well known from standard work in quantum field theory that 
different regularization procedures are not equivalent. For example, 
dimensional  
regularization and momentum space techniques are not completely equivalent 
as regards their treatment of the symmetries of the system. Since we have no 
freedom in choosing a regularization, it is necessary to accept and 
investigate the final results arising from the ansatz. The use of modified 
propagator to the field 
theoretic calculation is a more realistic approach towards the removal of the 
divergences in field theory based on the relevant physics, rather than a 
formalistic approach based on improper mathematical manipulations. In an 
earlier work \cite{paddy2}, it was shown that the physical parameters in this 
system like mass and coupling constant have additive finite terms, that 
are proportional to the powers of $L_P$ by calculating the effective potential 
for a self-interacting scalar field theory using the modified propagator. 

	In this paper, we evaluate the second order radiative corrections in 
QED using the modified propagator. The three corrections are to the vacuum 
polarization, electron self-energy, and vertex function. In the conventional 
QED calculations the divergent terms are absorbed into the physical parameters 
like mass, charge and spin. The modified propagator here again acts as a 
regulator to the ultra-violet divergences in the theory. The modified 
propagator introduces two kinds of regulators in the radiative correction 
calculations, which are logarithmic and power law. In the three radiative 
corrections the leading order power law regulator is ${\mathcal{O}}(L_P^2 m^2)$ 
and this is very small when compared to the other non-divergent terms in the 
conventional QED results. The three renormalization factors has the 
logarithmic corrections which is of the ${\mathcal{O}}(\ln(L_P m))$.

	 The following point needs to be stressed regarding the actual values 
of the corrections. In pure QED, the perturbative expansion is in a series in
$\alpha$ and corrections are of the order of $\alpha \simeq 10^{-2}$, 
$\alpha^2\simeq 10^{-4}$ etc in successive orders. The 
lowest order quantum gravitational corrections are, by and large, of order 
$L_P^2 m^2 \simeq 10^{-45}$. Since $\alpha^{22}\approx (L_P m)^2$, the first 22 order 
corrections of QED will dominate over the quantum gravitational corrections 
computed here! Much before this, electroweak corrections will start modifying 
the results. Thus the finite corrections computed here are only of conceptual 
significance --- providing the lowest order corrections {\it from quantum 
gravity} --- rather than of any operational significance. The only exception to 
this general situation is when quantum gravitational effects break a symmetry 
originally present in the theory, which --- as we shall see --- does happen.

 	In sections II and III, we evaluate the second order radiative 
corrections in QED, and we discuss the results and present other applications 
in Sec. IV.  			
\clearpage

\section{Vacuum Polarization and Electron Self Energy}
\noindent
\subsection{Vacuum Polarization}
\noindent
	The interaction of the photon field with electron field modifies the 
free photon propagator. The photon propagator of momentum $q$, with the one 
loop radiation correction included, is given by 
$$ i D'^F_{\mu \nu}(q) = i D^F_{\mu \nu} + i D^F_{\mu \rho} \frac{i \Pi^{\rho 
\sigma}(q)}{4 \pi} i D^F_{\sigma \nu}, $$
\noindent where $\Pi^{\mu \nu}$ is the vacuum polarization tensor, and 
$D^F_{\mu \nu}$ is the free photon propagator. The free photon propagator, 
$D^F_{\mu\nu}$, in the above relation is the conventional QED propagator. We 
concentrate here on the corrections to $D_{\mu \nu}$ arising from the 
modification of $\Pi_{\mu\nu}$ rather than from direct modification of 
$D_{\mu\nu}$ due to our ansatz. Using the Feynman rules of QED in the momentum 
space, we can write the vacuum polarization tensor as 
\begin{eqnarray} 
\Pi_{\mu\nu}(q)\!& =&\! 16\pi i e^2 \! \int\!\! \frac{d^4k}{(2 \pi )^3}\! 
\l(k_{\mu}(k-q)_{\nu}\!+\!(k - q)_{\mu}k_{\nu}\!
-\! g_{\mu \nu}(k^2\! - \!qk\! -\! m^2) \r) \nonumber \\
& \times & G_F^P(k) G_F^P(k - q). 
{\label {mainvacc}}
\end{eqnarray}
\noindent Substituting for the propagator from Eqn.(\ref {maineq}), and 
following similar calculations as in conventional QED, the vacuum polarization 
tensor can be separated into gauge invariant and gauge non-invariant part \cite{greiner}. The resultant gauge invariant part is given by
\b{\label {gaugeinvar}}
\Pi_{\mu\nu}^1(q^2) = -\frac{4 e^2}{\pi} 
\l(q_{\nu}q_{\mu} - g_{\mu \nu} q^2 \r)
\int_0^1 dz z(1 - z) K_0(\xi),
\e
\noindent where 
$$ \xi^2 = L_P^2 \frac{m^2 - q^2 z(1 -z)}{z(1 -z)}. $$
\noindent The series expansion of $K_0(z)$, about the origin, is given by
\b {\label {k0series}}
K_0(z) =  -\gamma-\ln(z/2) - \frac{z^2}{4}\l[ 1 - \gamma - 
\ln(z/2)\r] + \cdots.
\e
\noindent Hence, Eqn.(\ref{gaugeinvar}) takes the form 
\begin{eqnarray}
\Pi_{\mu\nu}^1(q^2)&=& 
\l(q_{\nu}q_{\mu} - g_{\mu \nu} q^2 \r) \Bigg[(Z_3 - 1) - \frac{2 e^2}{\pi}
A_1 - \frac{e^2 L_P^2}{\pi} \Bigg(\frac{q^2}{2}A_1 \nonumber \\
&+&\frac{1}{2}\l(1-\gamma-\ln(L_P m/2)\r)\l(m^2-\frac{q^2}{6}\r)\Bigg)\Bigg]
{\label {mainvacc1}},
\end{eqnarray}
\noindent where
\b
A_1=\int_0^1 dz~ z(1-z) \ln\l(\frac{m^2}{m^2 - q^2 z(1-z)}\r) 
\e
\noindent to the lowest order of $K_0(\xi)$. The term $A_1$ is the familiar 
conventional QED non-divergent term and the remaining terms (of the order 
${\mathcal{O}}(L_P^2)$) are the leading order quantum gravitational power law 
corrections/regulators to the conventional QED terms. The 
contribution of the quantum gravitational corrections  to the vacuum 
polarization is extremely small as compared to the conventional non-divergent 
QED term, i.e. they are of the order $10^{-45}$ which is much smaller compared 
to the next order radiative corrections of QED. But as we can notice the 
quantum gravitational corrections to the conventional QED non-divergent terms 
become important when $L_P q \simeq 1$. In this high momentum transfer limit, 
the propagation effects of the virtual photons will probe the small scale   
quantum gravitational effects and the corrections will be of 
the same order as the conventional QED non-divergent terms. The charge  
renormalization factor, $Z_3$, which is divergent in 
the usual QED calculations is now finite and is given by 
\b
Z_3 - 1 =  \frac{2 e^2 \ln (L_P m/2)}{\pi} \l[\frac{1}{3} + \frac{6\gamma - 5}
{18 \ln (L_P m/2)} + {\mathcal{O}}(L_P^2 m^2)\r]. 
\e
\noindent The estimate of this factor comes out to be
\b
Z_3 - 1 = \frac{2 e^2} {3 \pi} \ln(L_P m) \simeq -0.1.
\e
\noindent In the earlier work \cite{paddy2}, authors have calculated charge 
renormalization factor using Effective action approach, which is a 
non-perturbative technique. In this approach, they calculated the effect of 
the classical electro-magnetic background on the quantum charged scalar fields, 
propagating in the flat space-time, using the above modified propagator. The 
charge renormalization factor they obtain using the effective action approach 
is of the form 
$ Z^P = \frac{q^2}{6 \pi} K_0(2 L_P m).$
On expanding $K_0(z)$ using Eqn.(\ref{k0series}), their estimate of 
the charge renormalization factor was also of the order of $0.1$.
	
	In the standard QED calculations, there does arise a gauge 
non-invariant part of the vacuum polarization tensor which is divergent and 
is renormalized 
to zero(for instance, see Hatfield \cite{hatfield}). Even though, this is a 
standard result and can be found in textbooks we have given the relevant 
steps in the Appendix for the sake of completeness. By retracing 
the steps given in Appendix, we now obtain a finite quantity to the gauge 
non-invariant part of the vacuum polarization tensor using the modified 
propagator. The gauge non-invariant part comes out to be
\begin{eqnarray} 
\Pi_{\mu \nu}^2(q)&=&\frac{ i L_P^2 e_0^2}{(4 \pi)^2} g_{\mu \nu}
\int_0^{\infty} \frac{dz}{z} \int_0^{\infty} \frac{dz'}{z'} \frac{1}
{(z + z')^2} \exp\l(\frac{i L_P^2}{4} (z^{-1} + z'^{-1})\r) \nonumber \\
&\times & \exp i\l(\frac{q^2 z z'}{z + z'} - m_0 (z + z') \r)
{\label {gnon1}}.
\end{eqnarray} 
\noindent In Eqn.(\ref{gnon1}), transforming the variables $z$ and $z'$ to a 
new set of variables by the relation $z = 1/t$ and $z' = 1/t'$ leads to  
\begin{eqnarray}
\Pi_{\mu \nu}^2(q)&=&\frac{ i L_P^2 e_0^2}{(4 \pi)^2} g_{\mu \nu}
\int_0^{\infty}t~ dt \int_0^{\infty}t'~ dt' \frac{1}{(t + t')^2}
\exp\l(\frac{i L_P^2}{4} (t + t')\r) \nonumber \\
&\times &\exp i\l(\frac{q^2}{t + t'} - m_0 (t^{-1} + t'^{-1}) \r)
{\label {gnoni}}.
\end{eqnarray} 
\noindent Using the standard identity,

$$ 1 = \int_0^{\infty} \frac{d\beta}{\beta} \delta\l(1 - \beta(z + z')\r)$$
\noindent and scaling $t_i \rightarrow t_i/\beta$, as in the conventional QED 
calculations, we get
\b{\label{maingnon}}
\Pi_{\mu \nu}^2(q) = -\frac{2 i e_0^2}{(4 \pi)^2} g_{\mu \nu} L_P^6 \int_0^1 
t(1 - t) \frac{K_2(\xi)}{\xi^2}~dt,
\e
where 
$$\xi^2 = L_P^2 \l(\frac{m_0^2}{t(1 - t)} - q^2)\r.$$
\noindent The expansion of $K_2(z)$ near the origin is given by
$$K_2(z) = \frac{2}{z^2} - \frac{1}{2} + \l( \frac{3 - 4\gamma}{32} + 
\frac{\ln(2) - \ln(z)}{8} \r) z^2 + \cdots.$$
\noindent Substituting the series expansion of $K_2(\xi)$ in Eqn.(\ref{maingnon}), we obtain
\begin{eqnarray}
\Pi_{\mu \nu}^2(q)&=&g_{\mu \nu} \frac{-2 i e_0^2}{(4 \pi)^2} \Bigg[2 L_P^2 
\int_0^1 dt \frac{(t - t^2)^3}{(m_0^2 - q^2 t(1 -t))^2} \nonumber \\
&-& \frac{L_P^4}{2} \int_0^1 dt \frac{(t - t^2)^2}{(m_0^2 - q^2 t(1 -t))} 
+ \cdots \Bigg]
{\label{lastnon}}.
\end{eqnarray}
\noindent This clearly shows that the the gauge non-invariant part is nonzero 
for $L_P \neq 0$ and vanishes as $L_P \rightarrow 0$. While the term which 
breaks the gauge invariance is small to be of operational significance, it 
does have certain conceptual importance. The following points need to be noted 
regarding this result:
\vskip 10pt
\noindent (i) Mathematically speaking, this result arises from the fact that 
our ansatz is equivalent to a momentum space regularization procedure in 
conventional QED. It is known that, momentum space regularization, in
contrast to dimensional regularization, can lead to gauge breaking terms.
Usually, this is considered as an argument in favor of dimensional 
regularization. In our approach, of course, we have no choice and the result
arises automatically. In fact, it is very likely that any quantum
gravitational cutoff will appear like a momentum space regulator and will
break the gauge symmetry.
\vskip 10pt
\noindent(ii) Previously, Ohanian \cite{ohanian1} had obtained a gauge breaking 
term using a gravitationally smeared propagator. There is however one  
vital difference between our result and the one obtained by him. Note that, in  
our approach, the propagator reduces to that of conventional field theory when
$L_P\to 0$. If the procedure is to be consistent, the gauge breaking 
term should vanish when the limit of $L_P\to 0$ is taken. This is true 
as regards our result in Eqn.(\ref{lastnon}) showing that this is indeed a 
quantum gravitational effect. However, Ohanian \cite{ohanian1} obtains 
a gauge breaking term which does not vanish in the corresponding limit. This 
suggests that, our approach does allow a consistent interpretation of the 
results.
\vskip 10pt	
\noindent (iii)	It is the extra term $L_P^2 (z + z')^{-2}/(4 z z')$ which 
breaks the gauge invariance of the electro-magnetic field. This extra factor 
can be associated to the current in the charge conservation relation and hence, 
implying that the charge conservation is no more valid. There have been -- 
more drastic! -- attempts in the literature to break even the Lorentz 
invariance by introducing a coupling of the photon to charged scalar field, 
through gravitational couplings to the photon, etc \cite{magnetic}. 
The basic aim in the process of breaking the conformal symmetry of the 
electro-magnetic field is allowing for the possibility of generating large 
scale magnetic fields within inflationary scenarios. Most of these studies 
have used ad-hoc interaction potential to break the conformal invariance. The 
breaking of gauge invariance of the electro-magnetic field in our case is from 
a much more deeper ``principle of path integral duality". The connection to 
the cosmological seed magnetic field is still under investigation. 

\subsection{Electron Self Energy}
\noindent	
	The interaction of the electron field with photon field modifies the 
free electron propagator. The electron propagator of momentum $p$, with the 
one-loop correction included, is given by
$$ iS'_F(p) = iS_F(p) + iS_F(p)(-i\Sigma(p))iS_F(p), $$
where $\Sigma(p)$ (a 4-spinor) is the self-energy function, and $S_F(p)$ is 
the free electron propagator. Using the Feynman rules in the momentum space, 
we get
\b
\Sigma(p) =  -4 \pi i e^2 \int \frac{d^4k}{(2 \pi)^4} \gamma^{\mu}(\gamma^
{\nu} p_{\nu} - \gamma^{\nu} k_{\nu} + m) \gamma_{\mu} G_F^P(k) G_F^P(p - k).
\e
\noindent Substituting the propagator from the Eqn.(\ref{maineq}), the above 
equation reduces to
\b {\label{elecself}}
\Sigma(p) = \frac{e^2}{4 \pi^2} \int_0^1 dz \l(2 m_0 - \gamma^{\mu} p_{\mu} z 
\r) K_0(\xi),
\e
\noindent  where $$\xi^2 =  \frac{L_p^2}{z (1 -z)} \l[m_0^2 z - p^2 z (1 -z) 
\r]. $$
\noindent On expanding $K_0(\xi)$ using Eqn.(\ref{k0series}) we obtain, the 
lower order terms corresponding to the conventional QED results and the higher 
order terms as contributions of the quantum fluctuations of the 
space-time to the self energy of electron. The electron wave function 
renormalization, and the shift in the mass are obtained by recasting the 
interacting field propagator to look like the free field propagator i.e. we 
set
$$Z_2(\gamma^{\mu}p_{\mu} - m_0 - \Sigma(p)) = \gamma^{\mu}p_{\mu} - m + 
\mathrm{finite\ terms}.$$

	The electron wave function renormalization factor is obtained by 
equating terms proportional to $p$ in the above equation. This results in 
\b
Z_2^{-1} - 1 = \frac{e^2}{8 \pi^2} \l[ \gamma + \ln 2 - \ln(L_p m_0) + 
{\mathcal{O}}(1) \r]. 
\e
\noindent[Note that we have neglected the higher order terms of $K_0(\xi)$ as 
these have the dependence as $L_P^2 m_0^2$ whose contributions are negligible].
 The last term in the above expression is a finite quantity and is of the order one and it has the dependence of the electron momentum. The shift in the mass is given by
\b
\frac{\delta m}{m_0} = - \frac{e^2}{8 \pi^2} \l[3\ln (L_P m_0/2) + 3 \gamma +
{\mathcal{O}}(1) \r]
\e
\noindent The estimate of the scale factor $Z_2$ comes out to be
\b
Z_2^{-1} - 1 = - \frac{\alpha}{\pi} \ln{m_0 L_P} \simeq 0.1,
\e
\noindent and the fractional shift in the mass is
\b
\frac{\delta m}{m} \simeq - \frac{\alpha}{\pi} \ln{m_0 L_P} \simeq 0.1.
\e    
\noindent The usual infrared divergences is ignored in the calculation of the 
renormalization factor $Z_2$. Here again, we see that both the mass 
renormalization factor and the mass shift are also of the same order as the 
charge renormalization factor i.e. $0.1$.

\section{Vertex Correction and Anomalous magnetic moment}
\noindent
\subsection{Vertex Correction}
\noindent
For a free Dirac field, the current density is defined as 
$$J^{\mu} = \bar{\psi}\gamma^0\gamma^{\mu}\psi.$$
\noindent Thus, in the low energy QED processes the current transfer is 
related by $\gamma^{\mu}$, which is the low energy vertex function. The 
radiative corrections will modify the vertex, to the one-loop correction, as 
\b
-ie\Lambda_{\mu} = -ie\gamma_{\mu} -ie\Gamma_{\mu},
\e 
\noindent where, $\Gamma_{\mu}$ is the vertex function. Using the Feynman 
rules, we obtain
\begin{eqnarray}
\Gamma_{\mu}(p',p)&\!\!\!=&\!\!\!(-i e_0)^2 \int\!\! \frac{d^4 k}{(2 \pi)^4} 
\l[ \gamma_{\mu} (\gamma_{\rho} p'^{\rho} - \gamma_{\rho} k^{\rho} + m_0) 
\gamma_{\mu} (\gamma_{\sigma} p'^{\sigma} - \gamma_{\sigma} k^{\sigma} + 
m_0)\r]\nonumber \\
&\times & G_F^P(k) G_F^P(p' - k) G_F^P(p - k).
\end{eqnarray}
\noindent Substituting for the propagator from the Eqn. (\ref {maineq}), the 
above equation reduces to 
\begin{eqnarray}{\label{mainvertex}}
\Gamma_{\mu}(p', p)&\!\!\!=&\!\!\!(Z_1^{-1} - 1) \gamma_{\mu} + 
\frac{(-i e_0)^2}{(4 \pi)^2} \int_0^1 dz \int_0^{1 - z} dz' 
\frac{\xi}{T_1}  \\ \nonumber
&\times &\l[ 2 \gamma_{\mu} (p' - p)^2 (1 - z') (1 -z) - 
4 i m z (1 - z - z') (p' - p)^{\mu} \sigma_{\mu \nu}\r],
 \end{eqnarray}
\noindent where 
$$ \xi^2 =  L_P^2 T_1 T_2,~~T_1 = (z p' + z' p)^2 - z p'^2 - z' p^2 + (z + z') 
m_0^2, $$
\noindent and
$$T_2 = (1 - z - z')^{-1} + z^{-1} + z'^{-1}.$$

\noindent The vertex renormalization factor $Z_1$ is given by
\begin{eqnarray} {\label{vertex}}
Z_1^{-1} - 1&=&i \frac{(-i e_0)^2}{(4 \pi)^2} \int_0^1 dz \int_0^{1 - z} dz' 
\Big[ 4 K_0(\xi) + 2 m_0^2 \frac{\xi}{T_1} K_1(\xi) \nonumber \\
&\times & (-2 + 2(z + z') + (z + z')^2) \Big].
\end{eqnarray}
\noindent Substituting for $K_0(\xi)$ using Eqn.(\ref{k0series}), and using 
the series expansion of $K_1(\xi)$ as
\b{\label{k1series}}
K_1(z) = \frac{1}{z} + \frac{z}{2} \l( \ln(z/2) + \frac{2 \gamma - 1}{2} \r) + 
\frac{z^3}{16} \l(\ln(z/2) - \frac{5 - 4 \gamma}{4} \r),
\e
\noindent in the Eqn. (\ref{vertex}), the vertex normalization factor comes 
out to be
\begin{eqnarray}
Z_1^{-1} - 1 & = &\! i \frac{(-i e_0)^2}{(4 \pi)^2} \int_0^1\!\! dz 
\int_0^{1 - z}dz' \bigg[ -4 \ln(\xi/2) - 4 \gamma  + \frac{2 m_0^2}{T_1} 
\nonumber \\ 
&\times & (-2 + 2(z + z') + (z +z')^2)\bigg]  \\
& \cong & -\frac{\alpha}{\pi} \ln(m_0 L_P),
\end{eqnarray}
\noindent which is roughly of the order of $0.1$. The estimated renormalization
 factors $Z_1$ and $Z_3$ using the modified propagator are not equal unlike in 
conventional QED. 

\subsection{Anomalous Magnetic Moment} 
\noindent
	The triumph of QED has been the precision test of electron anomalous 
magnetic moment. The experimental value of the anomalous magnetic moment of an 
electron is in excellent agreement  with the predicted perturbative 
calculations up to the $4^{th}$ order to the $15^{th}$ decimal place. It is 
therefore of interest to compute the quantum gravitational corrections to the 
magnetic moment of an electron using the modified propagator.  

	 The vertex correction contribution to scattering of an electron in an 
external field is given by $\bar{u}(p')\Gamma^{\mu}(p',p)u(p)A_{\mu}^c(p'- p)$,
 where $\bar{u}$ and $u$ are the spinor wave-functions. Since, our interest is 
in calculating the radiative and the quantum gravitational corrections to the 
gyro-magnetic ratio of an electron, the term involving $\sigma_{\mu \nu}$ in 
the Eqn. (\ref{mainvertex}) is of our concern. The corresponding $\mathcal{M}$ 
matrix for this process is given by \cite{greiner},
\begin{eqnarray}
\mathcal{M}& = &\bar{u}(p')\gamma_{\mu}^M u(p)\\ \nonumber
&\!\!\! = &\!\!\!\! \bar{u}(p') 4 m_0 \frac{(i e_0)^2}{(4 \pi)^2} 
\int_0^1\!\!\!\! dz \! \int_0^{1 - z}\!\!\!\! dz' z(1 - z - z') 
(p' - p)^{\nu} \frac{\sigma_{\mu \nu} \xi}{T_1} K_1(\xi) u(p).
\end{eqnarray}
\noindent In the limit of small momentum transfer, $(p'-p)^2 \ll m_0^2$, one 
obtains by expanding $K_1(\xi)$ from the Eqn. (\ref{k1series}), we get
\begin{eqnarray}{\label{finalmag}}
\mathcal{M}&=& -\bar{u}(p') \frac{\alpha}{m \pi} (p'-p)^{\mu} \sigma_{\mu \nu} 
\int_0^1 dz \int_0^{1 - z} dz' \frac{z(1 - z - z')}{(z + z')^2} u(p) 
\nonumber \\
&-&\bar{u}(p') \frac{\alpha}{m \pi} \frac{L_P^2 m_0^2}{24} (p'-p)^{\mu} 
\sigma_{\mu \nu} \int_0^1 dz \int_0^{1 - z} dz' z(1 - z - z') T_2 \nonumber \\
& \times & \ln(L_P m_0^2 (z + z')^2 T_2/4) u(p). 
\end{eqnarray}
\noindent The first term in the above equation corresponds to the usual QED 
radiative vertex correction of the gyro-magnetic ratio (g) to the order 
$e_0^2$. The second term in the above equation is the quantum gravitational 
corrections to the gyro-magnetic ratio of an electron. The integral in the 
second term of the Eqn. (\ref{finalmag}) is convergent. The contribution of 
the quantum gravitational correction to the gyro-magnetic ratio to the first 
order in $\alpha$ is of the order $L_P^2 m^2 \simeq 10^{-45}$. Obviously, this 
is not of practical significance.

\section{Conclusions and Discussion}
 In this paper, we evaluated the quantum gravitational corrections(QGC) to 
three radiative corrections, in the first order of $\alpha$, in QED using the 
``principle of path integral duality". The modified propagator is able to 
remove all the divergences which usually crop up in the conventional QED 
calculations. The main features of the modified propagator in QED are as 
follows:
\vskip 10pt
\noindent (a) The three renormalization factors($Z_1$, $Z_2$, and $Z_3$), and 
the mass shift are all of the same order, ${\mathcal{O}}(\ln(m L_P)) \simeq 
0.1$. The charge renormalization factor $Z_3$ using the non-perturbative
 methods in scalar QED is also of order $0.1$ \cite{paddy2}. The 
renormalization factors $Z_1$ and $Z_3$ are different in our case as opposed 
to the conventional QED calculations.
\vskip 10pt 
\noindent (b) The modified propagator makes the gauge non-invariant part of the 
vacuum polarization tensor to be non-zero for $L_P \neq 0$ vanishes and in the 
limit $L_P \rightarrow 0$. This breaking of the gauge symmetry is also related 
to also the difference between the two renormalization factors. We have 
briefly indicated the possible effect of this breaking of gauge invariance, to 
the generation of large scale magnetic fields within inflationary scenarios
\vskip 10pt
\noindent (c) The contribution of the QGC to the 
gyro-magnetic ratio is very small i.e. of the order of $L_P^2 m^2$ $ \simeq 
10^{-45}$. The contribution of the quantum gravitational correction to the 
vacuum polarization and the electron self energy is also of the same order. 

\nonumsection{Acknowledgments}
\noindent
S.S. is being supported by the Council of Scientific and Industrial Research, 
India.

\appendix{: Evaluation of Gauge Non-invariant part}
\label{sec:app1}

\noindent
For the sake of completeness, we outline the essential steps leading to the 
gauge non-invariant part of the vacuum polarization tensor in standard QED and 
how it is regularized to zero. The vacuum polarization tensor in the 
conventional QED calculations in $2n$ dimensions is given by
\begin{eqnarray}{\label{app1}}
\Pi^{\mu\nu}_{reg}(q)&=&(e \mu^{2-n})^2 \int\frac{d^{2n}k}{(2 \pi)^{2n}} 
\frac{i}{k^2 - m^2 + i \epsilon} \frac{i}{(k - q)^2 - m^2 + i\epsilon} 
\\ \nonumber
&\times & 2^n \l(k^{\mu} (k - q)^{\nu} + k^{\nu} (k - q)^{\mu} - 
((k^2 - m^2) - k\cdot q)g^{\mu\nu}\r),
\end{eqnarray}
\noindent where $\mu$ has the dimension of mass. Using the integral 
representation of the propagator, i.e.
\b
\frac{i}{k^2 - m^2 + i\epsilon} = \int_0^{\infty} dz \exp\l(iz\l(k^2 - m^2 + 
i\epsilon\r)\r)
\e
\noindent in Eqn.(\ref{app1}) and completing the squares in the exponential, 
we obtain
\begin{eqnarray}
\Pi^{\mu\nu}_{reg}(q)&=&(e \mu^{2-n})^2 \int_0^{\infty}\!\!
dz_1 dz_2 \int \frac{d^{2n}k}{(2 \pi)^{2n}} \\ \nonumber
&\times & 2^n \l(k^{\mu} (k - q)^{\nu} + k^{\nu} (k - q)^{\mu} 
- ((k^2 - m^2) - k\cdot q)g^{\mu\nu}\r) \\ \nonumber
&\times&\!\!\!\!\exp\l(i(z_1 + z_2)\l(k - \frac{z_2 q}
{z_1 + z_2}\r)^2\!\! + i \frac{z_1 z_2 q^2}{z_1 + z_2} - i(m^2 - 
i\epsilon)(z_1 + z_2)\r).
\end{eqnarray}
\noindent Shifting the variable of integration and using the relations,
$$ \int \frac{d^{2n}p}{(2 \pi)^{2n}} p^2 \exp(iap^2) = - \frac{n}{(4 \pi a)^n}
\frac{1}{a} \exp(i n \pi/2), $$
$$ \int \frac{d^{2n}p}{(2 \pi)^{2n}} p^{\mu} p^{\nu} \exp(iap^2) = - 
\frac{g^{\mu\nu}}{(4 \pi a)^n} \frac{1}{2 a} \exp(i n \pi/2),$$
\noindent we get
\b {\label{conva}}
\frac{\Pi^{\mu\nu}_{reg}(q)}{(e \mu^{2 - n})^2}=\frac{\exp(in\pi/2)}{(4\pi)^n}
\l[ g_{\mu\nu} Q(q^2, m^2) + (q^{\mu}q^{\nu} - g^{\mu\nu}q^2) P(q^2, m^2)\r],
\e
\noindent where
\begin{eqnarray}
\!\!\!\!Q(q^2, m^2)&\!\!\!\!\!\!=&\!\!\!\!\!\!\int_0^{\infty}\!\!\!\!\!
dz_1\!\! 
\int_0^{\infty}\!\! \!\!\!dz_2 \frac{2^n} {(z_1 + z_2)^n} \l(\frac{n - 1}
{(z_1 + z_2)} + i m^2 - i q^2 \frac{z_1 z_2}{(z_1 + z_2)^2}\r) \nonumber \\  
&\times &\exp\l(iq^2 \frac{z_1 z_2}{z_1 + z_2} - i(m^2 -i\epsilon)
(z_1 + z_2)\r), 
\end{eqnarray}
\noindent and
\begin{eqnarray}
P(q^2, m^2)&=&-i 2^{n + 1}\int_0^{\infty}~dz_1 \int_0^{\infty}~dz_2 
\frac{z_1 z_2}{(z_1 + z_2)^{n + 2}} \nonumber \\
&\times &\exp\l(iq^2 \frac{z_1 z_2}{z_1 + z_2} - i(m^2 -i\epsilon)
(z_1 + z_2)\r). 
\end{eqnarray}
\noindent Global gauge invariance of the action leads to current 
conservation, $\pa_{\mu}j^{\mu} = 0$. This implies 
$q_{\mu}\Pi^{\mu\nu}(q) = 0$. The factor proportional to $g^{\mu\nu}$ is the 
gauge non-invariant part of the vacuum polarization tensor and does not seem 
to satisfy the gauge invariance condition. 

	We summarize here the standard argument [see Hatfield \cite{hatfield}] 
to show that the first term, $Q(q^2, m^2)$, in Eqn.(\ref{conva}) 
can be shown to vanish and hence the regularization preserves the gauge 
symmetry.  Consider only the first term of the gauge non-invariant part 
and define 

\b
I(q^2, m^2) \equiv  \int_0^{\infty} dz_1 dz_2 \frac{1}{(z_1 + z_2)^{n + 1}} 
\exp\l(iq^2 \frac{z_1 z_2}{z_1 + z_2} - i m^2(z_1 + z_2)\r).
\e
\noindent Rescaling the integration variables, $z_1$ and $z_2$, in the above 
equation by $\beta$, i.e. $z_1 \rightarrow \beta z_1$ and $z_2 \rightarrow 
\beta z_2$, we obtain 
\begin{eqnarray}
I(\beta, q^2, m^2)& =& \frac{1}{\beta^{n - 1}} \int_0^{\infty} dz_1 dz_2 
\frac{1}{(z_1 + z_2)^{n + 1}} \nonumber \\
&\times &\exp\l(i\beta\l(q^2 \frac{z_1 z_2}{z_1 + z_2} - m^2(z_1 + z_2)\r)\r).
\end{eqnarray}
\noindent The quantity $I(\beta, q^2, m^2)$ is of course independent of the 
integration variables $z_1$ and $z_2$, and hence is also independent of the
parameter($\beta$), i.e. $\pa I(\beta, q^2, m^2)/\pa\beta$ $= 0$. 
Differentiating the above expression w.r.t $\beta$ and regrouping terms, we 
get
\begin{eqnarray}
0=-\beta \frac{\pa I}{\pa\beta}&\!\!\!\!=&\!\!\!\!\!\int_0^{\infty}\!\!\!\! 
dz_1 dz_2 \frac{\beta^{2- n}} {(z_1 + z_2)^n}\l(\frac{n - 
1}{(z_1 + z_2)} + i m^2 - i q^2 \frac{z_1 z_2}{(z_1 + z_2)^2}\r)\nonumber \\ 
&\times &\exp\l(i\beta\l(q^2 \frac{z_1 z_2}{z_1 + z_2} - m^2(z_1 + z_2)\r)\r).
\end{eqnarray}
\noindent Rescaling the variables $z_1$ and $z_2$ by $\beta^{-1}$, i.e. $z_i 
\rightarrow z_i/\beta$, in the above expression, we get
\begin{eqnarray}
\!\!\!\!\!\!\!\!\! 0 = -\beta \frac{\pa I}{\pa\beta}&\!\!\!\!=&
\!\!\!\!\!\int_0^{\infty}\!\!\!\!
dz_1 dz_2 \frac{2^n}{(z_1 + z_2)^n} \l(\frac{n - 1}{(z_1 + z_2)} + i m^2 - 
i q^2 \frac{z_1 z_2}{(z_1 + z_2)^2}\r)\nonumber \\
&\times &\!\!\exp\l(i\l(q^2 \frac{z_1 z_2}{z_1 + z_2} - m^2(z_1 + z_2)\r)\r)
=Q(q^2, m^2).
\end{eqnarray}
\noindent Thus, the gauge non-invariant part of the vacuum polarization 
tensor in the standard QED vanishes. Hence, $q_{\mu}\Pi^{\mu\nu}(q) = 0$. 

	In the rest of this appendix, we retrace the above steps for the 
modified propagator. The gauge non-invariant part of the vacuum polarization 
tensor gets modified to the form,
\b {\label{modvar}}
\Pi_{\mu\nu}^{(2)}(q^2, L_P)=(e \mu^{2 - n})^2 \frac{\exp(in\pi/2)}{(4\pi)^n} 
g_{\mu\nu} \Pi_0^{(2)}(q^2, L_P),
\e
\noindent where
\begin{eqnarray}
\Pi_0^{(2)}(q^2, L_P)&\!\!\!\!=&\!\!\!\!\!\int_0^{\infty}\!\!\!\!dz_1 
dz_2 \frac{2^n}{(z_1 + z_2)^n} \l[\frac{n - 1}{(z_1 + z_2)} + i m^2 - i q^2 
\frac{z_1 z_2}{(z_1 + z_2)^2}\r] \\ \nonumber 
&\times & \exp\l[iq^2 \frac{z_1 z_2}{z_1 + z_2} - i(m^2 -i\epsilon)(z_1 + 
z_2) + \frac{i L_P^2}{4}(z_1^{-1} + z_2^{-1})\r],
\end{eqnarray}
\noindent when the modified propagator is used. We now show that the gauge 
non-invariant part is a finite quantity and is of the order $L_P^2$. 
Here again, we consider the first term of the gauge non-invariant part and 
define
\begin{eqnarray}
I(q^2, m^2, L_P)&\equiv & \int_0^{\infty} dz_1 dz_2 \frac{1}{(z_1 + z_2)^
{n + 1}} \exp\l(iq^2 \frac{z_1 z_2}{z_1 + z_2} - i m^2(z_1 + z_2)\r) 
\nonumber \\ 
& \times & \exp\l(\frac{i L_P^2}{4}(z_1^{-1} + z_2^{-1})\r).
\end{eqnarray}  
\noindent Rescaling the variables $z_1$ and $z_2$ by $\beta$, i.e 
$z_i \rightarrow \beta z_i$, and differentiating the resultant of the 
above expression w.r.t $\beta$ 
and regrouping terms, we obtain
\b {\label{modI}}
-\beta\frac{\pa I(\beta, q^2, m^2, L_P)}{\pa\beta} \equiv 
Q(\beta, q^2, m^2, L_P) - R(\beta, q^2, m^2, L_P),
\e
\noindent where
\begin{eqnarray}
Q(\beta, q^2, m^2, L_P)& = &\int_0^{\infty}\!\!\! 
dz_1 dz_2 \frac{\beta^{2 - n}} {(z_1 + z_2)^n} \\ \nonumber  
& \times &\l(\frac{n - 1}{(z_1 + z_2)} + i m^2 - i q^2 \frac{z_1 z_2}
{(z_1 + z_2)^2}\r) \\ \nonumber 
&\times&\!\!\! \exp\!\l[i\beta\l(q^2 \frac{z_1 z_2}{z_1 + z_2} - 
m^2(z_1 + z_2)\r) + \frac{i L_P^2}{4 \beta}(z_1^{-1} + z_2^{-1})\r],
 \end{eqnarray}
\noindent and
\begin{eqnarray}
R(\beta, q^2, m^2, L_P)& \equiv &\int_0^{\infty} dz_1 dz_2~ \frac{i L_P^2}
{4 z_1 z_2}~\frac{\beta^{-n}}{(z_1 + z_2)^n} \\ \nonumber
&\times &\!\!\!\exp\l(i\beta\l(q^2 \frac{z_1 z_2}{z_1 + z_2} - 
m^2(z_1 + z_2)\r) + \frac{i L_P^2}{4 \beta}(z_1^{-1} + z_2^{-1})\r).
\end{eqnarray}
\noindent The term $I(\beta, q^2, m^2, L_P)$ is independent of $\beta$ (by 
the arguments stated earlier in this appendix) and hence, the partial 
differential 
$\pa I/\pa\beta$ vanishes. Hence,
$$Q(\beta, q^2, m^2, L_P) = R(\beta, q^2, m^2, L_P). $$
\noindent Rescaling the variables $z_1$ and $z_2$ by 
$\beta^{-1}$, i.e. $z_i \rightarrow z_i/\beta$, in Eqn.(\ref{modI}), 
we get
\b {\label{last}}
Q_{rescaled}(q^2, m^2, L_P) = R_{rescaled}(q^2, m^2, L_P) = 
\Pi_0^{(2)}(q, L_P), 
\e
\noindent where
\begin{eqnarray}
R_{rescaled}&=&\frac{ i L_P^2}{4} \int_0^{\infty} \frac{dz_1}{z_1}  
\int_0^{\infty} \frac{dz_2}{z_2} \frac{1}{(z_1 + z_2)^n}
\exp\l(\frac{i L_P^2}{4} (z_1^{-1} + z_2^{-1})\r) \nonumber \\
& &\exp i\l(\frac{q^2 z_1 z_2}{z_1 + z_2} - m_0 (z_1 + z_2) \r),
\end{eqnarray}
\noindent and $\Pi_0^{(2)}(q, L_P)$ is the quantity proportional to the gauge non-invariant part of the vacuum polarization tensor defined in Eqn.(\ref{modvar}). Using our ansatz, we have shown that the gauge 
non-invariant part of the vacuum polarization tensor is a finite quantity and vanishes as $L_P \rightarrow 0$.

\nonumsection{References}


\noindent
\begin{thebibliography}{000}
\bibitem{ahluwalia}
D.~V.~Ahluwalia, {\sl Wave-Particle duality at the Planck scale: Freezing of 
neutrino oscillations} gr-qc/0002005; Achim Kempf, Gianpiero Manjaro and 
Robert~B.~Mann,\/ Phys.\ Rev.\ D\ {\bf 52}, 1108(1995).
\bibitem{paddy1}
T.~Padmanabhan,\/ Phys.\ Rev.\ Lett. {\bf 78}, 237(1992); \/ Phys.\ Rev.\ D\ 
{\bf 57}, 6206 (1998).
\bibitem{paddy2}
K.~Srinivasan, L.~Sriramkumar, and T.~Padmanabhan,\/ Phys.\ Rev.\ D\ {\bf 58}, 
044009 (1998).
\bibitem{ohanian1}
H.~C.~Ohanian,\/ Phys.\ Rev.\ D\ {\bf 55}, 5140 (1997).
\bibitem{greiner}
W.~Greiner, and J.~Reinhardt, {\sl Quantum ElectroDynamics} (Springer-Verlag, 
Berlin, Heidelberg, 1994).
\bibitem{hatfield}
B.~Hatfield, {\sl Quantum Field Theory of Point Particles and Strings} 
(Addison-Wesley Publishing Company, California, 1992), page 392.
\bibitem{magnetic}
M.~S.~Turner, and L.~M.~Widrow,\/ Phys.\ Rev.\ D\ {\bf 37}, 2743 (1988); 
B.~Ratra,\/ Ap. J. Lett. {\bf 391}, L1 (1992); O.~Bertolami, and D.~F.~Mota, 
{\it Primordial Magnetic Fields via Spontaneous Breaking of Lorentz 
Invariance}, gr-qc/9811087
\end{thebibliography}
\end{document}